\documentclass[prb,preprint,endfloats]{revtex4}

\def\nle{\ \raise.3ex\hbox{$<$}\kern-0.8em\lower.7ex\hbox{$\sim$}\ }
\def\nge{\ \raise.3ex\hbox{$>$}\kern-0.8em\lower.7ex\hbox{$\sim$}\ }

\usepackage{bm}
\usepackage{graphics,epsf}
\usepackage{array}
\usepackage{amsmath}
\usepackage{amssymb}

\begin{document} 

\title{Brownian Dynamics Studies on DNA Gel Electrophoresis. I.\\
Numerical Method and Quasi-Periodic Behavior of Elongation-Contraction Motions}
\author{Ryuzo Azuma}
\altaffiliation[Present address: ]{Genomic Sciences Center, RIKEN,
1-7-22 Suehiro, Tsurumi, Yokohama, Kanagawa, 230-0045, Japan}
\affiliation{Institute for Solid State Physics, The university of Tokyo
5-1-5 Kashiwanoha, Kashiwa, Chiba 277-8581, Japan}
\author{Hajime Takayama}
\affiliation{Institute for Solid State Physics, The university of Tokyo
5-1-5 Kashiwanoha, Kashiwa, Chiba 277-8581, Japan}
\date{\today}

%\wideabs{
\begin{abstract}
Dynamics of individual DNA undergoing constant field gel electrophoresis
 (CFGE) is studied by a Brownian dynamics (BD) simulation method which
 we have developed. The method simulates electrophoresis of DNA in a 3
 dimensional (3D) space by a chain of electrolyte beads
 of hard spheres. Under the constraint that the separation of each pair
 of bonded beads is restricted to be less than a certain fixed value, as
 well as with the excluded volume effect, the Langevin equation of
 motion for the beads is solved  by means of the Lagrangian multiplier
 method. The resultant  mobilities, $\mu$, as a function of 
 the electric field coincide satisfactorily with the
 corresponding experimental results, once the time, the length and the
 field of the simulation are properly scaled. In relatively strong
 fields quasi-periodic behavior is found in the chain dynamics, and is
 examined through the
 time evolution of the radius of the longer principal axis, $R_l(t)$. It
 is found that the mean width of a peak in $R_l(t)$, or a period of one
 elongation-contraction process of the  chain, is proportional to the
 number of beads in the chain, $M$, while the mean period between two
 such adjacent peaks is proportional to $M^0$ for large $M$. These
 results, combined with the observation that the chain moves to the
 field direction by the distance proportional to $M$ in each
 elongation-contraction motion, yield $\mu \propto M^0$. This explains
 why CFGE cannot separate DNA according to their size $L\ (\propto M)$
 for large $L$.
\end{abstract}
%}
\maketitle

\section{Introduction}

Gel electrophoresis is a most widely used technique to separate DNA 
according to their length $L$ under an electric field.~\cite{viovy00}
The constant
field method is quite efficient for DNA of small sizes. It is well
known, however, that the separation does not work well for DNA of sizes
longer than a certain value which depends on the field and the ``pore''
diameter of the gel.~\cite{nord91} The non-constant field
methods such as the pulsed-field gel electrophoresis~\cite{schcan}
have been empirically invented and they have remarkably improved the
shortcoming. Nevertheless, the understandings on these gel
electrophoresis methods, even on the constant field method, still 
remain unsatisfactory from the view point of statistical physics of
polymers. 

By constant field gel electrophoresis (CFGE), DNA molecules with the
radius of inertia, $R_{\rm I}$, smaller than the pore diameter of gel, 
$a_{\rm gel}$, are {\it sieved} by gel.~\cite{rodb70,chra71} In this
sieving regime DNA with the smaller $R_{\rm I}$ migrate the faster
through the gel. In the opposite case with $R_{\rm I}\gtrsim 
a_{\rm gel}$, DNA is considered to move through a ``primitive 
chain''~\cite{DoEd86} which is determined by stochastic movement of its
front end in pores of the gel under the
field.~\cite{lump82,lump85,slat86} This picture, denoted as the biased
reptation model, successfully describes various features of CFGE, such
as the $L$-dependence of the migration mobility, $\mu$, observed
experimentally~\cite{nanc85} in sufficiently weak fields. More
recently, the model of the biased reptation with fluctuations has been
proposed~\cite{sdv95} which well describes the observed field dependence
of $\mu$ in moderate fields.~\cite{viovy00}
%For DNA with over a few kbp (kilobase-pairs), 
When the field is further increased, the field-dependence of $\mu$
begins to be saturated.  
In this regime, interesting behavior of an individual DNA has been 
observed: ``hernias''-like configurations by the computer 
simulations~\cite{deut88,deut89} and ``quasi-periodic'' behavior, i.e.,
``elongation-contraction'' motion in a cyclic way by the
real-time fluorescent microscopy experiments.~\cite{masu93,oana94}
Indeed, the latter experimental method combined
with optical or magnetic tweezers for a molecule reveals that the
elasticity of DNA is originated from the entropy associated with its
static conformations.~\cite{smith92,perk95,stric96,aust97} However,
dynamic properties of DNA under CFGE are not described in a unified way
by this picture alone. 

In order to get further insights into the problem mentioned above, 
we have developed a Brownian dynamics (BD) method which
simulates electrophoresis of DNA in a 3 dimensional (3D) space.  
The method has been extensively applied to the CFGE, in particular, the
elongation-contraction motion of the chain under CFGE, the results of
which we will discuss in the present series of papers. In the present
paper I our numerical method is explained and global aspects, such as
the mean mobility $\mu$ and time evolutions of the velocity of the
center of mass, $v_{\rm G}(t)$, and of the radius of the longer
principal axis, $R_l(t)$, are investigated. In the accompanying 
paper II~\cite{ourII} we analyze dynamics of `defects' introduced by de 
Gennes~\cite{deGe71} in a chain under CFGE. Finally in paper
III~\cite{ourIII} we
will examine conformational changes of the coarse-grained chain under
CFGE in details, and will argue that the field-parallel component of its  
elongation-contraction motion is essentially understood as a
deterministic dynamics of an elastic string in a 1D space with an
obstacle, thereby an origin of the elasticity will be shown to be the
conformational entropy of the chain.

In the Brownian dynamics (BD) simulations extensively performed in the
present series of studies, DNA is modeled as a chain of spherical
electrolyte beads with a constant radius $1/2$ in a 3D space filled with
a solvent. Each bead interacts with other beads of the chain and
spatially-fixed beads constituting gel by hard-core repulsive interactions. 
It also interacts with its two bonded beads by a non-linear elastic
force which keeps the distance between the two to be less than a
threshold value which we put $\sqrt{2}$. To solve the Langevin equation 
of motion under these constraints as well as under the field and random
forces, we  
employ a BD algorithm with the Lagrangian multiplier method. 
It is similar to the one by Deutsch and Madden,~\cite{deut88,deut89}
but is essentially different from their method in the following respects.
In their method, the time pitch to solve the equation is adjusted so as 
to keep a fixed distance between bonded beads under the tensile forces
which are evaluated by the Lagrangian multiplier method. In our method,
on the other hand, the time pitch is fixed at a certain value, which we 
regard as the mean period of random forces. Within each time step we 
evaluate paths of all beads affected by the constraints, i.e., by
hard-core interactions and extremely nonlinear elastic interactions
between a pair of beads. These effects are reread as the supposed 
constraining forces with the appropriate Lagrangian multipliers in an
iterative way. The resultant constraining forces thus specified yield a
new set of bead positions without any violation to the constraints. 
To our knowledge the present BD analysis is the first extensive
simulation of DNA gel electrophoresis in a 3D space based on the
Langevin equation involving purely mechanical, microscopic forces which 
are supposed to act on monomers of DNA.

The purposes of the present paper I are to explain our BD method and to
demonstrate its results on some global properties of DNA under CFGE. The
resultant mobility $\mu$ as a function of the field turns out to agree
well with
those observed experimentally once a set of three parameters, $l^*$,
$\tau^*$ and $E^*$, are properly specified. They relate respectively the
scales of length, time and field strength of the simulation and the
experiments. 

In the BD simulation under relatively large fields quasi-periodic time
evolution is observed in $R_l(t)$ (see Fig.~\ref{fig:Fig-8} below),
as well as in $v_{\rm G}(t)$, as has been observed 
experimentally.~\cite{masu93,oana94} Each local maximum of
$R_l(t)$ at $t_{\rm max}^n$ with $n=1,2,...$ corresponds to a 
conformation that a chain just gets rid of trapping due to gel. The local
minimum at $t_{\rm min:l}^n$ preceding this maximum is then considered
to be the instance that, being trapped by gel, the chain starts to
elongate from a certain coiled conformation,  while at time 
$t_{\rm min:r}^n$ of the local minimum subsequent to $t_{\rm max}^n$ the
chain retrieves another coiled state. We establish a method to properly
classify a whole time sequence of $R_l(t)$ into peak parts, from 
$t_{\rm min:l}^n$ to $t_{\rm min:r}^n$ and the rest. We call the former 
parts {\it deterministic} ranges and the latter {\it stochastic} ones.
We then investigate statistics of periods of the deterministic
ranges, $D_n \equiv t_{\rm min:r}^n - t_{\rm min:l}^n$, and those of the
stochastic ranges, $(1/\lambda)_n \equiv 
t_{\rm min:l}^{n+1} - t_{\rm min:r}^n$. It is found that \{$D_n$\} obey
a Gaussian distribution with the mean $D \equiv \overline{D_n} \propto
M$ and the width $\Delta \propto D$, while  \{$(1/\lambda)_n$\} do a
Poisson distribution with the mean $1/\lambda \equiv 
\overline{(1/\lambda)_n} \propto M^0$ for large $M$, where $M$ is the
number of beads of the chain and the overline stands for the averages
over $n$, i.e., peak structures. These results, combined with
the observation that in each elongation-contraction motion the chain
moves to the field direction by the distance proportional to $M$, yield 
$\mu \propto M^0$ and explain why CFGE cannot separate DNA according
to their sizes $L\ (\propto M)$ for large $L$.

The organization of the present paper is as follows. In the next section
we explain our BD method. The mobility $\mu$ of the chain
obtained by the BD method is presented and compared with the
experimental results in Sec.~\ref{sec:mocuge}. The method of analysis of
peak structures in $R_l(t)$ is explained, and statistical nature of the
{\it deterministic} and {\it stochastic} ranges is discussed in 
Sec.~\ref{sec:ecmw}. The last section is devoted to further discussions
of our results.

\section{The Brownian Dynamics Method}\label{sec:method}\label{sec:bdm}

A chain of spherical beads in a continuous 3D space, shown schematically
in Fig.~\ref{fig:Fig-5}, is investigated by a new type 
of Brownian dynamics (BD) calculation. The hard core interaction is
assumed in order for centers of any pair of beads not to come closer
than 1 (the excluded volume effect), where
the radius of beads is set $1/2$. Furthermore distances between 
neighboring beads, $l$, are restricted to $l<\sqrt{2}$. We suppose that
this is realized by the extremely nonlinear elastic interaction
between the neighboring beads: it is infinite for $l>\sqrt{2}$ and
zero otherwise. 
These interactions assure the self-avoiding walk of the chain.

\begin{figure}
%\begin{center}
%\leavevmode\epsfxsize=80mm
%\epsfbox{sfge-bd/band.ps}
\caption[]{A snapshot of $M=10$ chain simulated in the BD method.
The spherical objects express beads with a radius $0.5$.
There is no overlap between any pair of beads. 
The bar-like objects are inserted only to visualize bonding 
of neighboring beads. 
}
\label{fig:Fig-5}
%\end{center}
\end{figure}

The BD method we adopt in the present work is to solve the following 
Langevin equation of motion for the chain,
\begin{equation}
\zeta \dot{\bm x}_i = q{\bm E}_{\rm b} 
+ \sum_\alpha \sum_j {\bm S}^\alpha_{ij} 
+ \sum_\alpha \left[ {\bm F}^\alpha_i-{\bm F}^\alpha_{i-1}\right] 
+ {\bm f}_i,
\label{eqn:lange}
\end{equation}
where ${\bm x}_i$ is the position of the center of the $i$-th bead, and
$\zeta$, $q$, and ${\bm E}_{\rm b}$ are the viscosity, 
the charge of a bead and 
the (bare) external field, respectively. 
The second and the third terms in r.h.s. are the constraining forces
required for the beads to satisfy the conditions,
\begin{eqnarray}
\|{\bm x}_{ij}\|=&\|{\bm x}_{j}-{\bm x}_i\|&>1\label{eqn:cexcl}\\
\|{\bm l}_i\|=&\|{\bm x}_{i+1}-{\bm x}_i\|&<\sqrt{2}.\label{eqn:cbond}
\end{eqnarray}
The vectors ${\bm S}^\alpha_{ij}$ and ${\bm F}^\alpha_i$ are 
respectively written as $\theta^\alpha_{ij}[{\bm x}_i-{\bm x}_j]$
and $\phi^\alpha_i[{\bm x}_i-{\bm x}_{i+1}]$ by introducing 
the Lagrangian multipliers $\theta^\alpha_{ij}$ and $\phi^\alpha_i$. 
The last term in eqn.~\ref{eqn:lange} stands for the random force from
a solvent. Its distribution is characterized by the following averages
\begin{eqnarray}
\langle {\bm f}_i \rangle &=& {\bm 0}\\
\langle {\bm f}_i(t){\bm f}_j(t')\rangle &=& 2\zeta k_{\rm B}T\Delta t\delta_{ij}\delta(t-t').\label{eqn:chi}
\end{eqnarray}
Here, $k_{\rm B}$, $T$, and $\Delta t$ are the Boltzmann constant,
the absolute temperature, and the average period of the random force,
respectively. 

To solve the one step evolution of eqn.~\ref{eqn:lange},
thereby the terms $\{{\bm S}^{\alpha}_{ij},{\bm F}^{\alpha}_i\}$ with the 
Lagrangian multipliers are determined, we examine the full Langevin equation 
with the inertia term and the forces arising from the hard core interaction 
and the nonlinear elastic interaction mentioned above.
We solve the equation in a time 
interval of $\Delta t > t > 0$, and take the overdamped limit, 
$m/\zeta\to 0$, where $m$ is the supposed mass of the bead.
The constraints 
of eqns.~\ref{eqn:cexcl} and \ref{eqn:cbond} are supposed to be satisfied 
by positions of the beads at $t=0$, denoted by $\{{\bm x}^0_i\}$. At this 
instance ${\bm f}_i$ and $q{\bm E}_{\rm b}\Delta t$ are added 
as impulsive forces with the integrated strength $\tilde{\bm f}_i$ where 
$\tilde{\bm f}_i={\bm f}_i+q{\bm E}_{\rm b}\Delta t$. When the
constraints are discarded,  
the solution in the overdamped limit is described by the relation between 
the displacement $\Delta {\bm x}_i$ in this time interval and the force as
\begin{equation}\label{eqn:pred}
	\Delta {\bm x}_i=\frac{1}{\zeta}\tilde{\bm f}_i,
\end{equation}
which is just what is obtained by simply integrating eqn.~\ref{eqn:lange} 
without the second and the third terms.

\begin{figure}
%\begin{center}
%\leavevmode\epsfysize=85mm
%\epsfbox{sfge-bd/Mons.ps}
\caption[]{The scattering processes and the associated constraining forces
 to fulfill the constraint of eqn.~\ref{eqn:cexcl} (a) and
 eqn.~\ref{eqn:cbond} (b).}
\label{fig:scattering}
%\end{center}
\end{figure}

In the case that positions of a pair of beads
${\bm x}^{01}_i\equiv {\bm x}^0_i +\Delta {\bm x}_i$ and 
${\bm x}^{01}_j\equiv {\bm x}^0_j+\Delta {\bm x}_j$
violate the constraint of eqn.~\ref{eqn:cexcl}, we solve the full
Langevin equation for this pair with the hard core scattering, which
can be either elastic or inelastic. We then obtain, again in the
overdamped limit, positions ${\bm x}^1_i$ and ${\bm x}^1_j$ at 
$t=\Delta t$ as shown in Fig.~\ref{fig:scattering}a in case of the
inelastic scattering. The result is reread as the consequence of the
constraining force,
${\bm S}^1_{ij}\equiv \theta^1_{ij}[{\bm x}^1_i-{\bm x}^1_j]$,
which is determined by the conditions 
${\bm x}^1_i-{\bm x}^0_i=(\tilde{\bm f}_i +{\bm S}^1_{ij})/\zeta$ and 
${\bm x}^1_j-{\bm x}^0_j=(\tilde{\bm f}_j -{\bm S}^1_{ij})/\zeta$ (note
that ${\bm S}^1_{ji}=-{\bm S}^1_{ij}$). 
This constraining force ${\bm S}^1_{ij}$ just corresponds to the one in
the second term of eqn.~\ref{eqn:lange}. In the case ${\bm x}^{01}_i$
and ${\bm x}^{01}_{i+1}$ violate eqn.~\ref{eqn:cbond}, the nonlinear
elastic force is supposed to reduce the distance 
$\|{\bm l}_i\|$ to $\sqrt{2}$. This effect is represented
by the constraining force ${\bm F}^1_i\ (=-{\bm F}^1_{i+1})$, which is  
added to the third term of eqn.~\ref{eqn:lange}. Its Lagrangian
multiplier $\phi^1_i$ is determined in such a way that
$\|{\bm x}^1_i-{\bm x}^1_{i+1}\| =\sqrt{2}$ with  
${\bm x}^1_i={\bm x}^0_i+{\bm F}^1_i\Delta t$ and
${\bm x}^1_{i+1}= {\bm x}^0_{i+1}+{\bm F}^1_{i+1}\Delta t$
(see Fig.~\ref{fig:scattering}b). If ${\bm x}^{01}_i$ involves violation
to more than one constraints of eqns.~\ref{eqn:cexcl} and/or
\ref{eqn:cbond}, the above-mentioned procedures are applied to evaluate
the constraining forces for each pair of beads with the common 
${\bm x}^{01}_i$. If, on the other hand, ${\bm x}^{01}_i$ does not
involve any violation to eqns.~\ref{eqn:cexcl} and \ref{eqn:cbond}, we
put ${\bm x}^1_i={\bm x}^{01}_i$.

In the configuration $\{{\bm x}^1_i\}$, it is generally expected that new
pairs of beads may violate the constraints since we have tried to remedy
the violations in $\{{\bm x}^{01}_i\}$ in a pairwise way. Then we repeat
the above-mentioned procedure, in which $\tilde{\bm f}_i$ is replaced by
$\tilde{\bm f}_i+{\bm C}^1_i$, where ${\bm C}^1_i=\sum_j{\bm S}^1_{ij}
+{\bm F}^1_i-{\bm F}^1_{i-1}$ is the constraining force
determined by the first procedure. If a configuration without any violation
to the constraints is obtained by the $\alpha_{\rm con}$ times repetition
of this procedure, we regard $\{{\bm x}^{\alpha_{\rm con}}_i\}$ as the 
solution of eqn.~\ref{eqn:lange} evolved from $\{{\bm x}^0_i\}$ by one unit
of time $\Delta t$. In this manner the multi-scattering processes
in the interval of $\Delta t$,
including those due to the nonlinear elastic interaction,
can be taken into account, and the corresponding forces
$\{{\bm S}^{\alpha}_{ij},{\bm F}^\alpha_i\}$ in eqn.~\ref{eqn:lange} are
determined. 

Although our BD method is based on the overdamped Langevin equation
with the Lagrangian multipliers, it distinctly differs from the one
due to Deutsch and Madden.~\cite{deut88,deut89} 
In the latter the equidistant condition
on all the pairs of beads is imposed, and the constraining tensile forces
associated with it are evaluated at the instance when random forces are
applied. Then the time interval $\Delta t$ is adjusted so that the
configuration at $t + \Delta t$ in fact obeys the equidistant
condition within accuracy
of $0.1\%$. In contrast, our BD method takes into account a similar, but
looser, constraint on the bond length as well as the excluded volume effect
of the beads by explicitly examining the supposed multiple scattering within
the interval $\Delta t$ which is fixed. Therefore there is no finer time
unit other than $\Delta t$ in solutions of our BD method. It is the mean
period of random forces $\{{\bm f}_i\}$.

In the present work we simulate dynamics of a chain in a continuous 3D
space confined into a cubic box with a volume $360^3$. 
The periodic boundary condition is imposed to each direction of the box. 
The gel is represented by a network of immobile bars
arranged so that they form a simple cubic lattice. 
%%as shown in Fig.~\ref{fig:Fig-6}. 
Each bar consists of tightly connected beads with a radius $1/2$,
and its interaction with a bead of the mobile chain is computed according
to the constraint of eqn.~\ref{eqn:cexcl}. 
We employed three lattices of gel with different lattice distances
$a_{\rm gel}=10$, $17$, and $20$, which are  considered to correspond
to the 'pore' diameter of agarose gel.~\cite{nord91} 
The lengths of chains, $M$, used in the BD calculation are $M=30$, $40$,
$60$, $80$, $160$, $240$, and $320$. 
We have performed typically $64$ runs whose initial configurations
are given independently. The direction of the
field is fixed at $(1,1,1)$. The time is measured in unit of $\Delta
t$, i.e., 1 BD step.

Here we want to emphasize that the present BD method correctly
reproduces the expected static and dynamic properties of a real (not
phantom) polymer in a 3D space under the vanishing field:
$R_I\propto M^{\nu_{\rm F}}$ with $\nu_{\rm F}\simeq 0.59 \pm 0.01$
and the Einstein relation $D_G\propto \mu/M$ are confirmed (see
Fig.~\ref{fig:mobility} in Sec.~\ref{sec:mocuge}). 
Here, $\nu_{\rm F}$, $R_I$, and $D_G$ are the Flory exponent, the radius
of inertia, and  the diffusion constant of the center of mass, respectively. 

The number of the multiple scattering, $\alpha_{\rm con}$ above mentioned,
is expected to be at most of the order of $M^2$. It has turned out that
$\alpha_{\rm con}$ becomes relatively larger when the chain is in an
extended configuration under gel electrophoresis. We quote here typical
figures for a chain with $M=80$: $\alpha_{\rm con}\simeq 4\times 10^1$,
$5\times 10^2$ for $E=0$, $0.032$, respectively. 

\section{Mobility}\label{sec:mocuge}

The most basic quantity that has been commonly measured in CFGE of DNA 
is the long time average of velocity divided by the electric field, or 
the mobility $\mu$. In the present work, we have implemented long 
time BD simulations with a highly efficient algorithm using a massive 
parallel computer and have eventually obtained $\mu$, hereafter denoted 
by $\mu_{\rm BD}$, with enough statistical precision for all the set of 
parameters investigated.  Here we only mention typical figures of 
computation to get $\mu_{\rm BD}$ of a $M=240$ chain for each run: more 
than 150 hours are consumed on a single processor of SGI2800 system to 
pass $10^6$ BD steps. In the present work the following values of the 
parameters are chosen: $\Delta t=1$, $\zeta= 10$ and $k_{\rm B}T=2$. 
The results are shown in Fig.~\ref{fig:mobility} with the experimental 
data due to Heller {\it et al}.~\cite{hdv94} Here and hereafter 
the value of field $E$ stands for $qE_{\rm b}$ of eqn.~\ref{eqn:lange}. 

The simulated $\mu_{\rm BD}$ at $E=0$ in Fig.~\ref{fig:mobility} is
evaluated  through the Einstein relation 
\begin{equation}
 {\mu_{\rm BD}^0 \over M} = D_{\rm G} \equiv 
\lim_{t\rightarrow \infty}{1 \over 6t}
\langle({\bm R}_{\rm G}(t)-{\bm R}_{\rm G}(0))^2\rangle|_{E=0},
\label{eqn:E-relation}
\end{equation}
where ${\bm R}_{\rm G}(t)$ is the position of the center of mass of the
chain. The diffusion constant $D_{\rm G}$ in the r.h.s. is evaluated by
the BD simulation in a vanishing field in the time range
$t=(5\sim10)\times10^6$. The resultant $\mu_{\rm BD}^0$
turns out to be compatible with the  value of $\lim_{E\rightarrow 0}
\mu_{\rm BD}$ as seen in the figure.  This confirms that the Einstein
relation in fact holds in the present BD simulation for the chain. 
  
\begin{figure}
%\begin{center}
%\leavevmode\epsfxsize=110mm
%\epsfbox{sfge-bd/T17-MU-new-3.ps}
\caption[]{\makebox[\textwidth][l]{Field dependence of mobilities.}}
\label{fig:mobility}
%\end{center}
\end{figure}

In order to compare the BD results with the experimentally observed
data, we have to specify proper conversion of values of the parameters
in the simulation and the experiment. Let us first consider the length
scale of a chain.  In the accompanying paper II we examine
in detail the embodiment of `defects' due to de Gennes~\cite{deGe71} in
our BD chain and obtain the mean distance $\langle n\rangle$ of adjacent
'defects' along the chain under $E=0$ as $\langle n\rangle \simeq 5.7$. 
This value, in unit of the mean distance between neighboring beads, 
$\langle l\rangle$, is almost independent of the set of parameters 
$a_{\rm gel}$ and $M$ we have examined. Regarding $\langle n\rangle$ as
an estimate of the `persistent length' in the BD simulation, 
${\cal P}_{\rm BD}$, we put ${\cal P}_{\rm BD}=c\langle n\rangle$, 
where $c$ is an adjustable parameter whose value is nearly equal to
unity (note that $\sqrt{2} \ge \langle l\rangle \ge 1$).  
The conversion factor of the length-scale $l^*$ is then determined by 
\begin{equation}
 {\cal P}_{\rm exp} = l^* {\cal P}_{\rm BD} 
= cl^* \langle n\rangle,
\label{eqn:L-scale}
\end{equation}
where ${\cal P}_{\rm exp}$ is the persistence length experimentally
measured. With ${\cal P}_{\rm exp} \sim$ 60nm at $T_{\rm exp} \sim$ 300K
taken from Ref.~\onlinecite{volk94}, we obtain $cl^* \sim$ 10nm. 
  
Another length scale of importance is the mean pore size of gel, which is 
rather difficult to be accurately specified even in the 
experiments.~\cite{viovy00} Here we simply suppose that DNA in $1\%$
agarose corresponds to a BD chain in the gel with $a_{\rm gel}=17$. With 
${\cal P}_{\rm BD}$ above fixed, this value of $a_{\rm gel}$ is about three
times larger than that of ${\cal P}_{\rm BD}$.

Next let us relate $E$ in the BD simulation to the experimental field, 
$E_{\rm exp}$, in V/cm. For this purpose we impose that the ratios 
$\Delta \varepsilon/k_{\rm B}T$ in the BD calculation and the experiment 
be identical, where $\Delta \varepsilon$ is the energy that a part of 
chain of a length equal to the persistence length ${\cal P}$ gains when 
it moves a distance of ${\cal P}$ under the electric field: 
\begin{equation}
\frac{\sigma E_{\rm exp}{\cal P}_{\rm exp}^2}{k_{\rm B}T_{\rm exp}}
=\frac{E{\cal P}^2_{\rm BD}}{k_{\rm B}T_{\rm BD}},
\end{equation}
or
\begin{equation} 
E_{\rm exp}=E^*E;\ \ 
E^*={k_{\rm B}T_{\rm exp}{\cal P}_{\rm BD}^2 
\over 2\sigma {\cal P}_{\rm exp}^2} 
\simeq 5\times 10^2c^{-2}\ {\rm V/cm}.
\label{eqn:rel-E}
\end{equation}
where $\sigma$ is the density of charge of DNA, and 
$k_{\rm B}T_{\rm BD}=2$ has been used. Quoting further the experimental
value of $\sigma \sim$ 0.6e$^-$/{\AA},~\cite{volk94} as well as 
$cl^*\sim$ 10nm at $T_{\rm exp} \sim$ 300K in eqn.~\ref{eqn:L-scale}, we
obtain the last expression for $E^*$ in the above equation.  
As for the ordinates of Fig.~\ref{fig:mobility} we may relate $\mu_{\rm BD}$ 
to $\mu_{\rm exp} 
[{\rm cm}^2/{\rm V}\cdot{\rm s}]$ by
\begin{equation} 
\mu_{\rm exp}=\mu^*\mu_{\rm BD};\ \ 
\mu^*={l^* \over \tau^* E^* }
\simeq 2 \times 10^{-9}{c \over \tau^*}\ [{\rm cm}^2/{\rm V}\cdot{\rm s}], 
\label{eq:rel-mu}
\end{equation}
where $\tau^*$[s] is the real time which corresponds to one discretized 
time step $\Delta t$ in the BD calculation with $\zeta=10$ (or 
$\Delta t/\zeta=0.1$). 

The two sets of data of $\mu$ in Fig.~\ref{fig:mobility} are drawn as
follows. The data $\mu_{\rm exp}$ are plotted directly
using the units indicated on the right ordinate and the upper abscissa,
whereas the units for $\mu_{\rm BD}$ are indicated on the left ordinate
and the lower abscissa with $c^2=1.60$ and $1.26$ respectively for
$M=240$ and $320$. The scales of the two ordinates are related by means of
$\tau^*=1.4\times10^{-6}$. We see from the figure that the field
dependence of $\mu_{\rm BD}$, particularly that of $M=240$, well coincide 
with that of $\mu_{\rm exp}$ of DNA with the corresponding length
we have specified, i.e., $4.3$ and $6.5$ kbp DNA for $M=240$ and $320$,
respectively. This is quite satisfactory if we take into account the
fact that only three adjustable parameters are used to draw the two sets
of data in Fig.~\ref{fig:mobility}. We therefore consider that the
present BD model catches up even quantitatively essences of DNA dynamics
under CFGE.

When we compare $\mu_{\rm BD}$ with $\mu_{\rm exp}$ further in detail,
however, there are some unsatisfactory features in the BD results. As
compared with $\mu_{\rm exp}$ of $6.5$ kbp DNA, an $E$-independent
branch of $\mu_{\rm BD}$ has not been ascertained in the weak field
limit for the $M=320$ chain. One possible reason for this may be 
the insufficient time of simulation (up to $t=10^6$ in finite fields) 
for this length of chain in weak fields. Thus, within the limited CPU
time, it is a rather hard task for the present BD simulation to judge
the field dependence of $\mu_{\rm BD}$, proportional to $E$ or $E^2$, in
weak fields.~\cite{herb87,holms90,dvs94} 
In the strong field regime, on the other hand, $\mu_{\rm BD}$ tend to
saturate, while, $\mu_{\rm exp}$ are still significantly increasing with
$E$. These different tendencies are seen more clearly if we compare 
$\mu_{\rm exp}$ measured in $10$V/cm with the corresponding 
$\mu_{\rm BD}$ (not seen). The reason of this discrepancy is not clear
at the moment.

We have also examined $\mu_{\rm BD}$ of the chains in gels with 
$a_{\rm gel}=20$ and $10$. If only the volume ratio of gel is
considered, they correspond roughly to $0.75\%$ and $3\%$ agaroses,
respectively. Qualitatively, $\mu_{\rm BD}$ becomes larger and its
$E$-dependence becomes 
weaker for the larger $a_{\rm gel}$ as expected, but these tendencies
are quantitatively weaker than those observed experimentally. The
origins of the discrepancy may be attributed to our model for gel, i.e.,
a perfectly rigid, regular jungle gym. 

\section{Elongation-Contraction Motion}\label{sec:ecmw}

\subsection{Motion of an individual chain}

In fields larger than a certain value depending on $M$ and 
$a_{\rm gel}$, chains in the present 3D BD simulation exhibit quasi-periodic 
behavior, i.e., they exhibit elongated and contracted shapes 
alternatively, as observed in the experiments~\cite{masu93,oana94,lars95} 
and in the previous 2D BD models.~\cite{deut88,matsu94,mas95p}
It is here emphasized that such quasi-periodic behavior has been
observed within the field range where we have discussed the
$E$-dependence of $\mu$ in Fig.~\ref{fig:mobility}. We show a typical
conformational change of a chain observed in one MD run of CFGE with the
parameters $M=240$, $E=0.032$, and $a_{\rm gel} = 17$ in 
Fig.~\ref{fig:varfigs-new}. Within a time window of the figure the 
elongation-contraction motion occurs three times at around $t=$2, 6 
and 11 $\times 10^5$.  
At ranges between them centered around $t=4$ and 8 $\times 10^5$ a 
chain is in a rather compact form. In this range it can happen, though 
not so frequently, that the front and rear ends exchange their roles as 
seen around $t=9\times 10^5$ in the figure. It is also pointed out here
that the front end of a chain moves in the field direction with almost a 
constant rate. 

\begin{figure}
%\begin{center}
%\leavevmode\epsfxsize=110mm
%\epsfbox{sfge-bd/v2j.ps}
\caption[]{A time evolution of projection of an $M=240$ chain.
$E=0.032$ $a_{\rm gel}=17$. Solid line stands for locus of $i=1$,
and dashed line for $i=240$. }
\label{fig:varfigs-new}
%\end{center}
\end{figure}

\subsection{Quasi-periodic behavior of $R_l(t)$ and 
$v_{\rm G}(t)$}\label{sec:quasi-p}

In Fig.~\ref{fig:Fig-8} we show the time evolution of the radius of 
longer principal axis, $R_l(t)$, 
and the velocity of the center of mass, $v_{\rm G}(t)$, in the MD run 
whose chain conformational change is shown in Fig.~\ref{fig:varfigs-new}.
The fluctuation in $R_l(t)$ is notable, showing
successive $\Lambda$-shaped peaks. We also note that just before the
maxima of $R_l(t)$, minima are observed in $v_{\rm G}(t)$.
These features are qualitatively in good agreement with those observed in
the experiment.~\cite{oana94} 

\begin{figure}
%\begin{center}
%\leavevmode\epsfxsize=80mm
%\epsfbox{sfge-bd/Fig-8.eps}
\caption[]{\makebox[\textwidth][l]{Time development of $R_l(t)$ and $v_{\rm G}$ under $E=0.032$.}}
\label{fig:Fig-8}
%\end{center}
\end{figure}

The time evolution of $R_l(t)$ and $v_{\rm G}(t)$ in
Fig.~\ref{fig:Fig-8} looks quasi-periodic. Actually it was found by the
experiment that the autocorrelation function of $R_l(t)$, 
$C_{\rm RR}(t)$, exhibits a damped oscillation.~\cite{oana94} This is
also the case for our present BD data. In Fig.~\ref{fig:Fig-19} 
$C_{\rm RR}(t)$ evaluated for various $M$ are presented. For 
$M\ge80$, $C_{\rm RR}(t)$ exhibit the first undershoot below the line of
$C_{\rm RR}(t)=0$, and even the first overshoot is clearly seen for 
$M\ge 160$. From these results we may conclude that, in chains with 
$M\ge 160$ under $E=0.032$, elongation-contraction motions occur
quasi-periodically. In the inset of Fig.~\ref{fig:Fig-19} we show nearly
linear dependence of the period of oscillation $\tau=4t_0$ on $M$, where
$t_0$ is defined as the time of the first intercept of $C_{\rm RR}(t)$
with the abscissa.~\cite{oana94} 

\begin{figure}
%\begin{center}
%\leavevmode\epsfxsize=80mm
%\epsfbox{sfge-bd/T17F0320.BLScor.ps}
\caption[]{The autocorrelation functions $C_{\rm RR}(t)$. 
The dependence of period $4t_0$ on $M$ is shown in the inset. }
\label{fig:Fig-19}
%\end{center}
\end{figure}

\subsection{Analyses on peak structures of $R_l(t)$}\label{sec:sop}

Oana {\it et al.}~\cite{oana94} argued that the steady-state time 
evolution of DNA under CFGE is classified into two types of behavior. One 
is the elongation-contraction motion which looks apparently deterministic.
We call the time branches in which such a motion is undergoing the 
{\it deterministic} ones (ranges centered at around $t=$2, 6 and 11 
$\times 10^5$ in Fig.~\ref{fig:varfigs-new}).  The other is what is 
observed between two deterministic branches. We call them {\it stochastic} 
branches. Oana {\it et al.} approximately described the quasi-periodic 
behavior of $R_l(t)$ in terms of a simplified function: peaks in $R_l(t)$ 
are approximated by $\Lambda$-shape branches whose widths are assumed to
obey a Gaussian distribution, and others by constants whose duration
times are assumed to obey a Poisson distribution. They calculated
$C_{\rm RR}(t)$ by means of this model function for various 
cases and remarked that when $\lambda D$ becomes larger than unity a
damped oscillation in $C_{\rm RR}(t)$ becomes prominent. 
Here $1/\lambda$ and $D$ stand for the mean period of the stochastic 
branches and the mean width of the deterministic ones, respectively. 

Following the classification due to Oana {\it et al.}, we analyze our BD
results by making use of an algorithm which appropriately picks up peaks
of $R_l(t)$. For this purpose we first pick up time sets 
$\{t^i_{\rm min:l}, t^i_{\rm max}, t^i_{\rm min:r}\}$, i.e., 
the times of preceding local minimum of the $i$-th maximum, 
the $i$-th maximum itself, and the subsequent minimum. Then we select out 
peaks whose height relative to their adjacent minima is larger than a 
certain threshold value. The latter is chosen in such a way that the sum
of periods of peak branches selected surmounts more than $50\%$ of the
total duration of the observation. This procedure 
divides a whole time sequence of $R_l(t)$ into the two types of 
branches whose numbers are equal to each other. It picks up ensemble of 
various peaks but not a limited numbers of peaks of special shapes. 
The number of peaks thus selected is more than a hundred for each set of 
parameters. 

\subsection{Statistics of deterministic and stochastic 
branches}\label{sec:statistics}

For the set of times $\{t^n_{\rm min:l}, t^n_{\rm max}, 
t^n_{\rm min:r}\}$ of peaks picked up by the procedure described above 
%in sec.~\ref{sec:sop}, 
we examine the distribution of peak widths (or 
periods of deterministic branch) $
D_n \equiv t_{\rm min:r}^n-t_{\rm min:l}^n$ and that of duration times 
between neighboring peaks (periods of stochastic branch) 
$1/\lambda_n\equiv t_{\rm min:r}^n-t_{\rm min:l}^{n-1}$. The results 
for $M=240, E=0.032$ and $a_{\rm gel}=17$ are shown in 
Fig.~\ref{fig:Fig-24}. They numerically confirm the basic assumptions 
of Oana {\it et al.}, i.e., $\{D_n\}$ nearly obey a Gaussian distribution, 
while $\{1/\lambda_n\}$ do a Poisson distribution. The mean and the variance 
of the former are given by $D=3.3\times10^5$ and $\Delta \simeq 0.2D$, 
respectively, while the mean width of the latter is given by $\lambda D 
\simeq 1.7$. The sum $D+1/\lambda$ should be equal to $\tau=4t_0$ 
evaluated from $C_{\rm RR}(t)$ in Sec.~\ref{sec:quasi-p}, which is verified 
within our numerical accuracy. The ratio $\lambda D\simeq 1.6$ fulfills
the condition for a damped oscillation to be observed in $C_{\rm RR}(t)$. 

\begin{figure}
%\begin{center}
%\leavevmode\epsfxsize=80mm
%\epsfbox{sfge-bd/bart.ps}
%\leavevmode\epsfxsize=80mm
%\epsfbox{sfge-bd/lambda.ps}
\caption[]{(a) Distributions of peak widths
${D}_n=t_{\rm min:r}^n-t_{\rm min:l}^n$ and
(b) duration between neighboring peak minima $1/\lambda_n=
t_{\rm min:r}^{n-1}-t_{\rm min:l}^n$
obtained for $M=240$, $E=0.032$, $a_{\rm gel}=17$. }
\label{fig:Fig-24}
%\end{center}
\end{figure}

The ratio $\lambda D$ is found to decrease below unity  for $M$ which 
is smaller than a certain value. In Fig.~\ref{fig:Fig-26}, $D$ and 
$1/\lambda$ are plotted against $M$ for $E=0.032$ and 
$a_{\rm gel}=17$. The results strongly suggest that $D\propto M^1$
and $1/\lambda\propto M^0$ for large $M$. In Fig.~\ref{fig:Fig-23}, the 
$M$-dependence of ratios $\lambda D$ and $\Delta/D$ are presented. As 
expected from the results shown in Fig.~\ref{fig:Fig-26}, $\lambda D$ is
an approximately linearly increasing function for large $M$. On the 
other hand, $\Delta/D$ is almost constant ($\sim 0.3$), indicating that 
the Gaussian distribution of $\{D_n\}$ is scaled by the mean $D$ alone. 
It should be noted that, for $M\nge 100$ where $\lambda D\nge 1$ holds, 
$C_{\rm RR}(t)$ clearly exhibits a damped oscillation as has been observed 
in Fig.~\ref{fig:Fig-19}. Also the result $4t_0 \propto M$ for 
$M \nge 100$ shown in the inset of Fig.~\ref{fig:Fig-19} is in 
accordance with the fact that $\lambda D$ is larger than unity for such 
$M$.  For sufficiently large $M$ the deterministic branches dominate 
a whole sequence of the time evolution, and so $4t_0 \simeq D \propto M$ 
holds. These results consistently indicate that there exists a 
crossover between the regimes with and without elongation-contraction 
motions at around $M\cong 100$ in CFGE with the parameters studied, i.e., 
$E=0.032$ and $a_{\rm gel}=17$.

\begin{figure}
%\begin{center}
%\leavevmode\epsfxsize=80mm
%\epsfbox{sfge-bd/T17F0320GTM.DPwkl1234.ps}
\caption[]{\makebox[\textwidth][l]{Dependence of $D$ and $1/\lambda$
on $M$. }}
\label{fig:Fig-26}
%\end{center}
\end{figure}

\begin{figure}
%\begin{center}
%\leavevmode\epsfxsize=80mm
%\epsfbox{sfge-bd/T17F0320GTM.DPwkl.ps}
\caption[]{\makebox[\textwidth][l]{Ratios $\lambda D$ and $\Delta/D$.}}
\label{fig:Fig-23}
%\end{center}
\end{figure}

We have also examined the field dependence of $D$ and $1/\lambda$ for a 
fixed $M (=240)$, the results of which are shown in 
Fig.~\ref{fig:E-dep-D}. In strong fields $E \nge 0.06$, $DE$ becomes
to be saturated, while 
$E/\lambda$ decreases with increasing $E$. As a consequence, $D\lambda$ 
is an increasing function of $E$, indicating that the deterministic 
branches, or in other words, the elongation-contraction motions survive 
and even dominate a whole evolution of the chain in sufficiently 
strong fields. 
\begin{figure}
%\begin{center}
%\leavevmode\epsfxsize=80mm
%\epsfbox{sfge-bd/B-L.ps}
\caption[]{\makebox[\textwidth][l]{Field dependence of $DE, E/\lambda$ and $D\lambda$ of $M=240$ chain.}}
\label{fig:E-dep-D}
%\end{center}
\end{figure}

Lastly, in Fig.~\ref{fig:Fig-13}, the averaged time evolution of
individual peaks of $R_l(t)$ are plotted for various $M$ and $E$. 
The abscissa and ordinate are normalized by the period $\tau\equiv 4t_0$
and the averaged peak height, respectively. It is seen that with these 
normalizations, the data with different $M$ and $E$ almost lie on a
universal curve during the period from $-0.6\tau$ to $0.4\tau$. The 
results obtained above such as $D \propto M, E^{-1}$, $\Delta 
\propto D (\propto M)$ can be derived from the scaling behavior of 
$R_l(t)$ in Fig.~\ref{fig:Fig-13}.  Such scaling behavior of $R_l(t)$ 
including the fact that the normalized peak exhibits an anti-symmetric 
shape is just what was observed experimentally.~\cite{oana94} 
It was mentioned in Ref.~\onlinecite{oana94} that the ratio of
slope of the $\Lambda$-shape immediately after the maximum to that
before it is approximately constant ($\sim 3$) for all
conditions they employed. The corresponding ratio of the BD result is 
$\sim 2$. The origin of this quantitative difference is not clear at the 
moment. 

\begin{figure}
%\begin{center}
%\leavevmode\epsfxsize=80mm
%\epsfbox{sfge-bd/periods.ps}
\caption[]{The averaged peak shape of $R_l(t)$. 
The abscissa is normalized by the period $\tau$
and the ordinate by the average of the peak height $\bar{R}_l(0)$. 
}
\label{fig:Fig-13}
%\end{center}
\end{figure}

\section{Discussions}\label{sec:discussions}

We have performed extensive simulations on the DNA constant-field gel 
electrophoresis (CFGE) by means of a new Brownian dynamics (BD) method 
which we have recently developed. Our BD method is based on a {\it 
generalized} bead-spring model for a polymer in a 3D space with rigid, 
regularly arrayed obstacles. By the word {\it `generalized'} we mean
that the `spring force' is not linear but extremely nonlinear: it is
infinite when the 
distance $l$ between two beads connected by the `spring' is larger than 
$\sqrt{2}$, while it is zero when $l < \sqrt{2}$. Also in the model we
strictly take into account of the excluded volume effects between any
pair of beads whose diameter is put unity. These forces are supposed to
be of a microscopic origin, i.e., molecular forces acting on monomers of
a polymer. It should be emphasized that there involves no characteristic
scale in these interactions except for the maximum distance between the
connected beads ($\sqrt{2}$) relative to the bead diameter (unity). 

In a vanishing field the viscosity $\zeta$ 
(or $\Delta t/\zeta$ with $\Delta t$ being the time unit of the BD 
dynamics) and the temperature $T$ in eqn.~\ref{eqn:lange} give rise to a
characteristic scale of the model. The two parameters specify the
viscous force and the random force from a solvent and are related to
each other by eqn.~\ref{eqn:chi}. As is discussed in the accompanying
paper II, the characteristic scale is the `persistence length' of the
chain. Actually, with $\zeta=10, \ \Delta t=1$ and 
$k_{\rm B}T_{\rm BD}=2$ the mean distance between two connected beads 
($cl^*$ in eqn.~\ref{eqn:L-scale}) turns out to be about one six-th of
the persistence length. Equating the latter to the persistence length of
real DNA, the mean bead distance is roughly estimated as 10nm. Thus a 
bead in our BD model represents a potion of DNA consisting of a few tens
of base-pairs.  We may call our BD model with these values of
parameters a bead-spring model of a semi-microscopic level. 
We will argue in paper III that the resultant chain dynamics, when it is 
coarse-grained in time and space, can be interpreted as dynamics of a
charged, elastic string, and that the elasticity is due to the 
conformational entropy of the original, semi-microscopic chain. 

A value of the electric field, $E=qE_b$ in eqn.~\ref{eqn:lange}, is
reasonably converted to that of the experimental CFGE on DNA as
discussed in Sec.~\ref{sec:mocuge}. By the procedure adopted there the
conversion of the field scale is essentially governed by that of the
length scale as expressed in eqn.~\ref{eqn:rel-E}. The parameter $c$ in 
eqn.~\ref{eqn:rel-E} has been adjusted for the chains with different sizes,
but within the range that the results are compatible with the condition 
$\sqrt{2} \ge \langle l\rangle \ge 1$ where $\langle l\rangle$ is the
mean distance of connected beads. Then the $E$-dependence of mobility 
$\mu_{\rm BD}$ has been shown to agree semi-quantitatively with that of
$\mu_{\rm exp}$ once absolute magnitudes of the simulational and
experimental $\mu$'s are adjusted at the vanishing field limit. This
procedure fixes the conversion of the time scales of the simulation and
the experiment. To our knowledge, such a semi-quantitative coincidence
between the simulation and the experiment has not been so far achieved.

We have also demonstrated that our BD simulation on CFGE in fields
stronger than a certain crossover value reproduces the quasi-periodic 
evolution of a chain whose characteristic aspects are quite similar to
those experimentally observed.
Actually, as proposed by Oana {\it et al.}~\cite{oana94} based on their
experimental observations, a whole sequence of time evolution of the
chain can be divided into two branches according to behavior of the
longer radius of the chain $R_l(t)$. One is the {\it deterministic}
branch  where $R_l(t)$ exhibits a distinct peak structure which
corresponds to the elongation-contraction motion of the chain, and the
other is the {\it stochastic} branch between two neighboring
deterministic branches where the chain exhibits a rather compact form. 

We have studied statistics of time durations of the two branches and
have found that, as assumed by Oana {\it et al.},~\cite{oana94} those of
the deterministic branch obey a Gaussian distribution with mean $D$ and
variance $\Delta$, while those of the stochastic branch do a Poisson 
distribution with the mean $1/\lambda$. Furthermore it is found that, in
a fixed field, $D\propto M^1,\ \Delta \propto D$ and 
$1/\lambda\propto M^0$ for large $M$. This means that the deterministic 
branches become more dominant in a whole time sequence, and that the 
quasi-periodicity in the chain dynamics becomes more distinct (see 
Fig.~\ref{fig:Fig-19}) for larger $M$. These results, combined with the
observation that the chain moves 
to the field direction by the distance proportional to $M$ in each
deterministic branch, yield $\mu \propto M^0$. This explains why CFGE 
cannot separate DNA according to their size $L\ (\propto M)$ for large 
$L$. We have also checked that, for a fixed $M$ which is sufficiently
large, the deterministic branches become dominant as $E$ is increased
at least within the field range of our BD simulation 
(see Fig.~\ref{fig:E-dep-D}). 

By the present BD simulation a crossover between the regimes with and
without elongation-contraction motions is expected to occur for a chain
with $M \simeq 100$ in $E=0.032$ and $a_{\rm gel}=17$. According to the
length and field conversions described in Sec.~\ref{sec:mocuge}, this
roughly corresponds to DNA of 2kbp in $1\%$ agarose in $E_{\rm exp}
\simeq 4{\rm V/cm}$. Experimently, on the other hand, the direct
measurement of the elongation-contraction motion by the fluorescent
microscopy has been rather limited; the shortest is for T7 DNA
(38kbp),~\cite{lars95} but mostly for T4 DNA (166kbp)~\cite{oana94} and
T2 DNA (164kbp).~\cite{lars95} But, there exist another experiment which 
strongly indicates the occurrence of the elongation-contraction motion,
i.e., the antiresonance of mobility in the filed inversion gel
electrophoresis. In such a experiment the antiresonance has
been clearly observed for DNA with 9.42kbp.~\cite{kdmo90} Therefore, our
criterion mentioned above is considered to be a reasonable estimate
for the elongation-contraction motion to occur. 

Lastly two further comments are in order. In the whole parameter ranges
we have examined, including the ones where the deterministic branches
sufficiently dominate the stochastic ones, so-called 'hernias'-like
configurations of a chain has been scarcely observed in the present BD 
simulation. This implies that they do not play a significant role in
determining the saturation of $\mu$ at least for chains with moderate
sizes we have studied.  Another interesting observation has been already
pointed out in Sec.~\ref{sec:quasi-p}: the front end of a chain moves in
the field direction with almost a constant rate 
(Fig.~\ref{fig:varfigs-new}). In relation to this, we note that the 
average of $v_{\rm G}(t)$ over the deterministic branches and that over
the stochastic branches are found to almost coincide with each other in
a whole range of field we have examined. We shall interpret these
results in paper III where the details of the chain dynamics in the
deterministic branches is discussed.

\begin{acknowledgments}

The authors wish to thank M. Doi and Y. Masubuchi for useful discussions
on their experimental and simulational results.
The computation in this work has been done using the facilities of 
the Supercomputer Center, Institute for Solid State Physics, 
University of Tokyo, and those of the Computer Center of University 
of Tokyo.

\end{acknowledgments}

\newpage
\bibliographystyle{apsrev}
\section*{References}
%\bibliography{bib/myrefs}
%\begin{thebibliography}{10}
%\end{thebibliography}

\newpage
\printfigures

\newpage
%\begin{figure}[t]
\begin{center}
\resizebox{!}{7mm}{FIG.~1}\\\vspace*{25mm}
\leavevmode\epsfxsize=80mm
\epsfbox{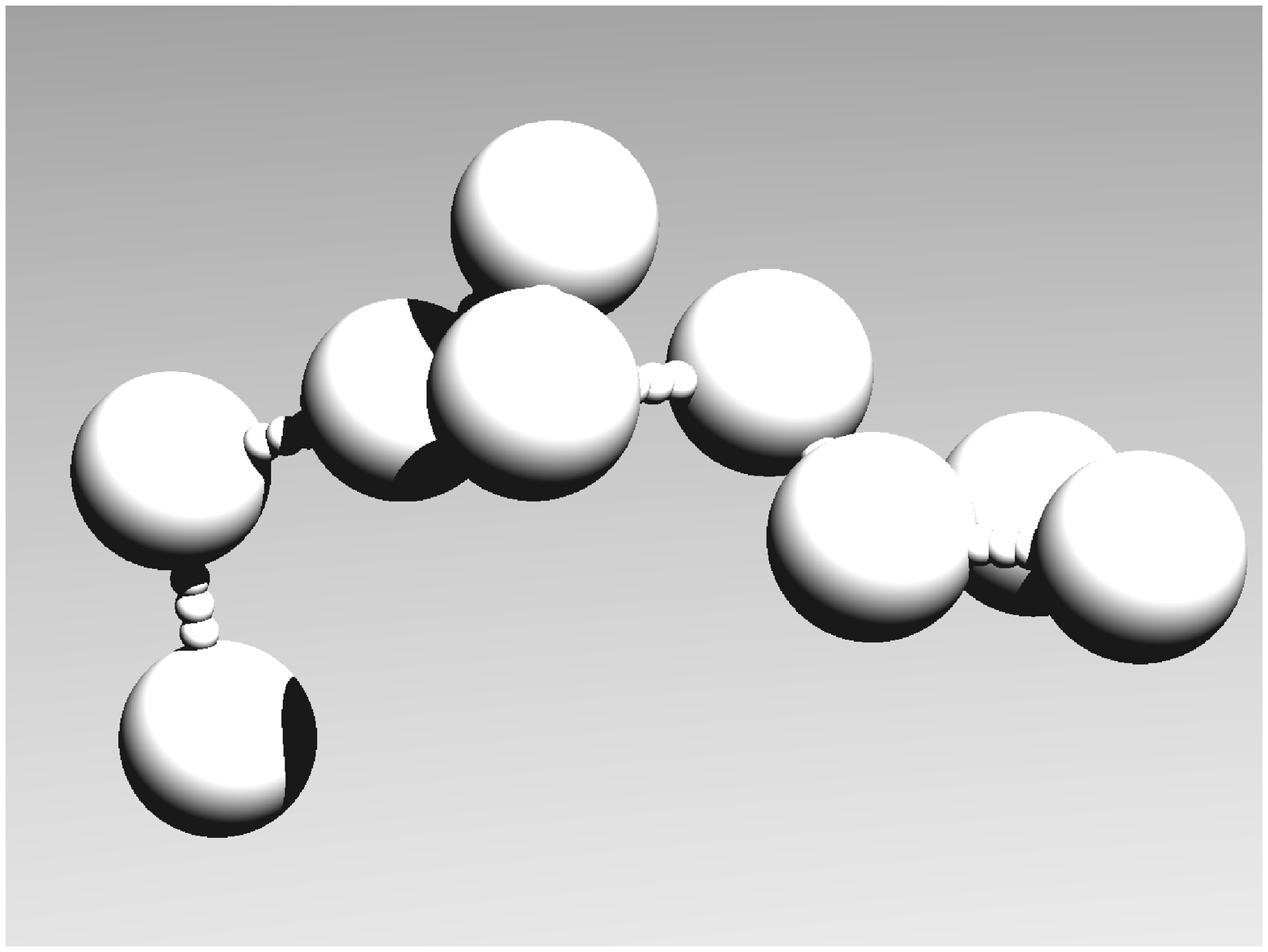}
\end{center}
%\end{figure}
\clearpage

\newpage
%\begin{figure}[t]
\begin{center}
\resizebox{!}{7mm}{FIG.~2}\\\vspace*{25mm}
\leavevmode\epsfysize=85mm
\epsfbox{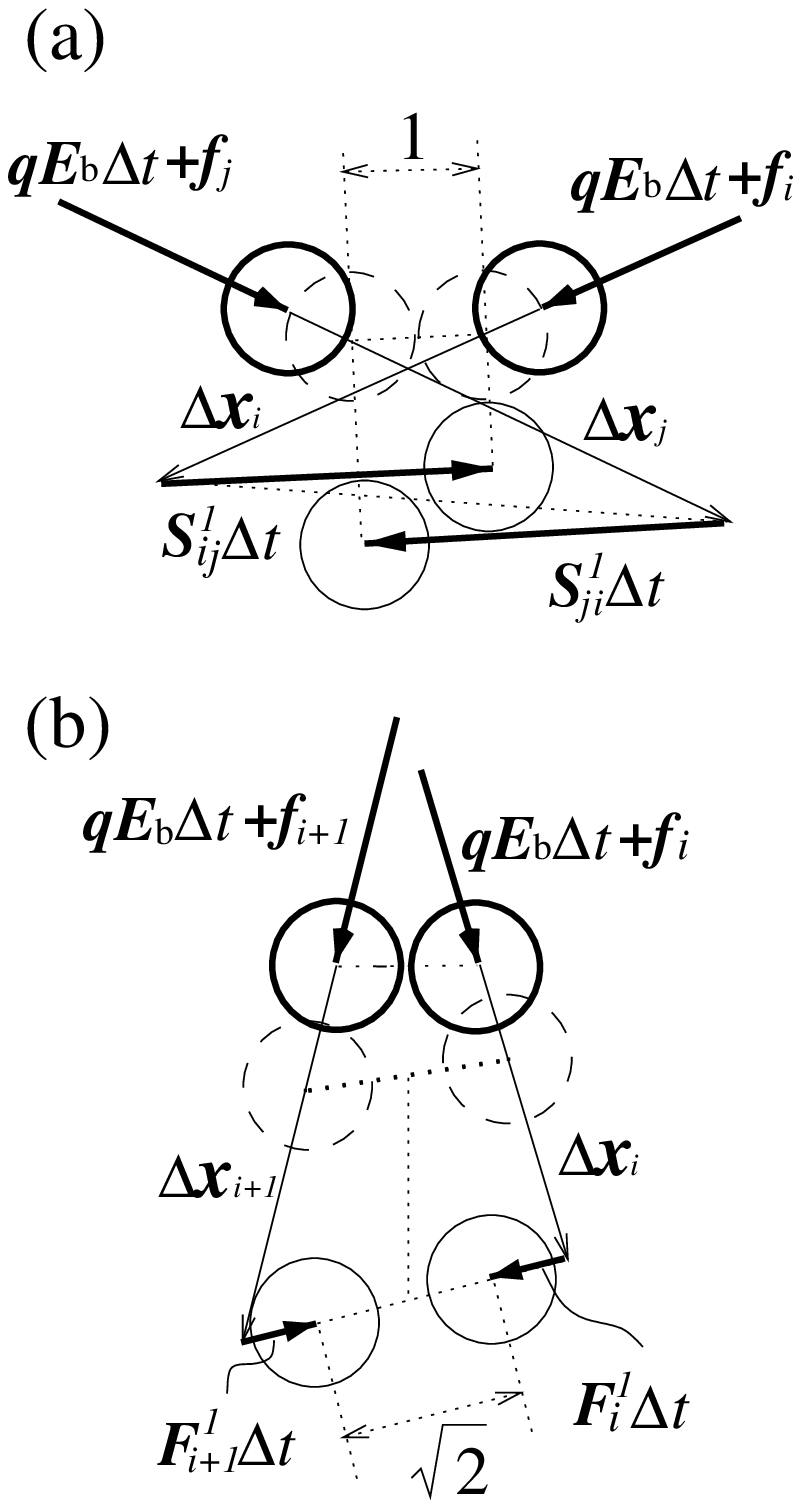}
\end{center}
%\end{figure}
\clearpage

\newpage
%\begin{figure}[t]
\begin{center}
\resizebox{!}{7mm}{FIG.~3}\\\vspace*{25mm}
\leavevmode\epsfxsize=80mm
\epsfbox{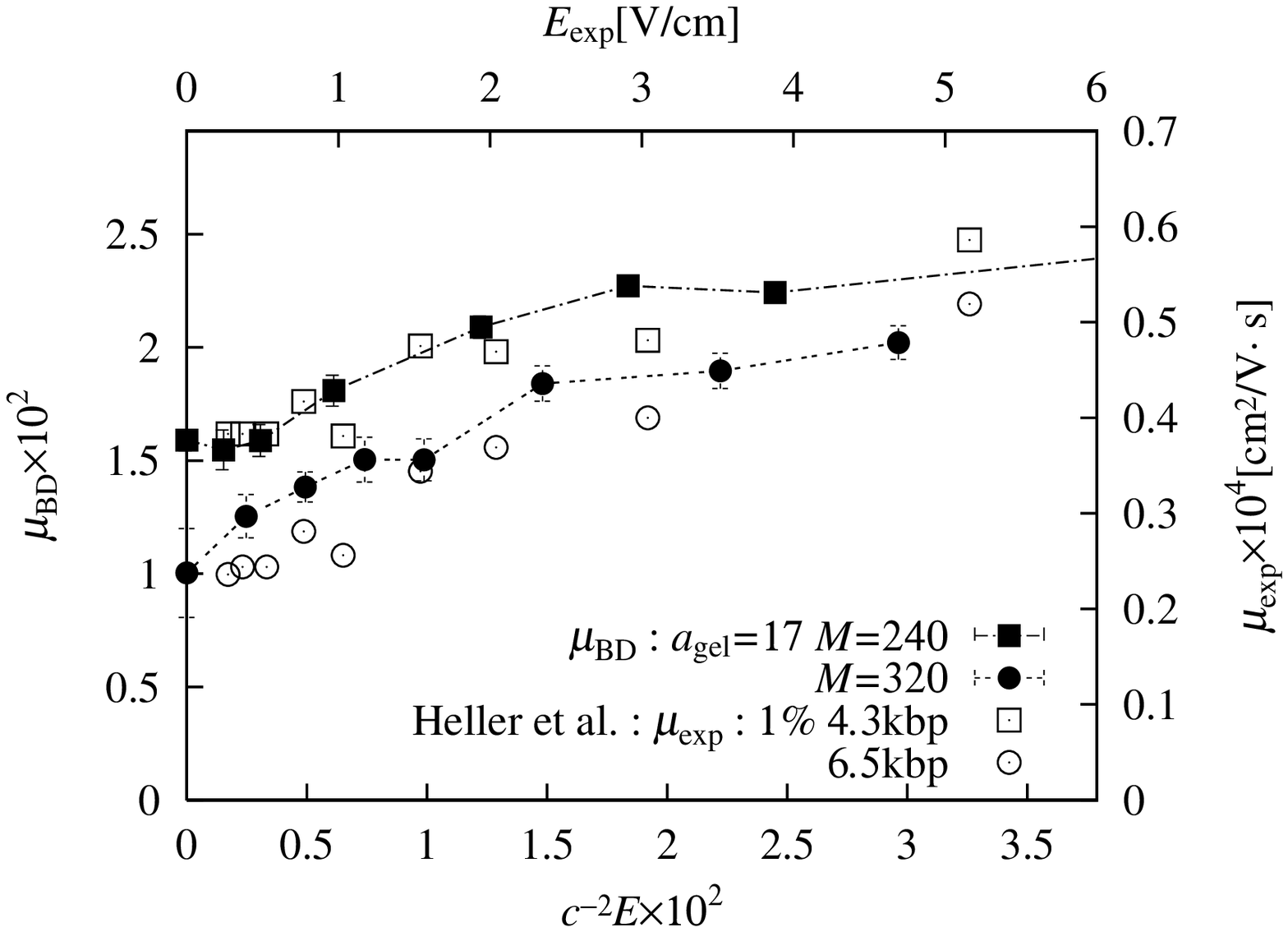}
\end{center}
%\end{figure}
\clearpage

\newpage
%\begin{figure}[t!]
\begin{center}
\resizebox{!}{7mm}{FIG.~4}\\\vspace*{25mm}
\leavevmode\epsfxsize=110mm
\epsfbox{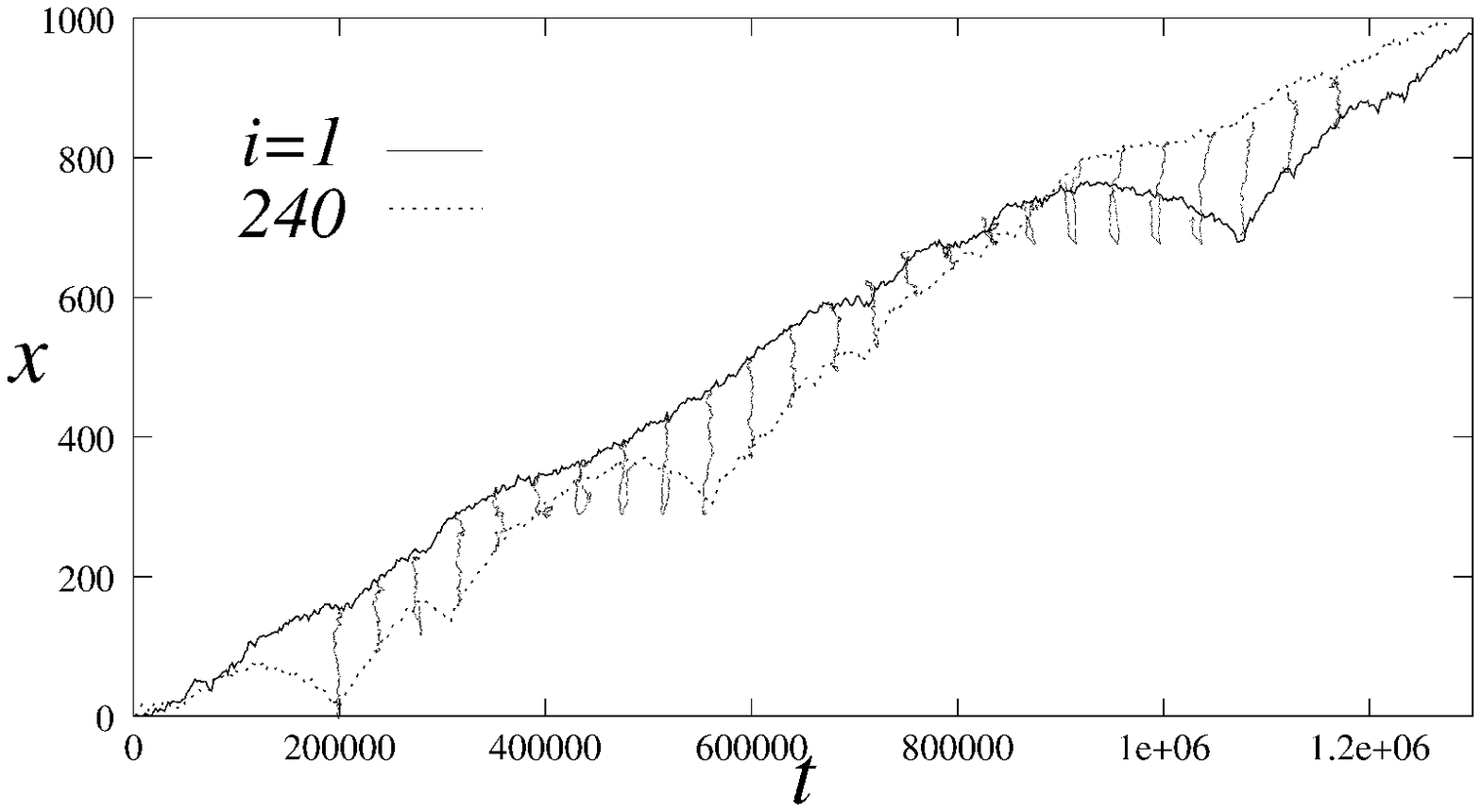}
\end{center}
%\end{figure}
\clearpage

\newpage
%\begin{figure}[t!]
\begin{center}
\resizebox{!}{7mm}{FIG.~5}\\\vspace*{25mm}
\leavevmode\epsfxsize=80mm
\epsfbox{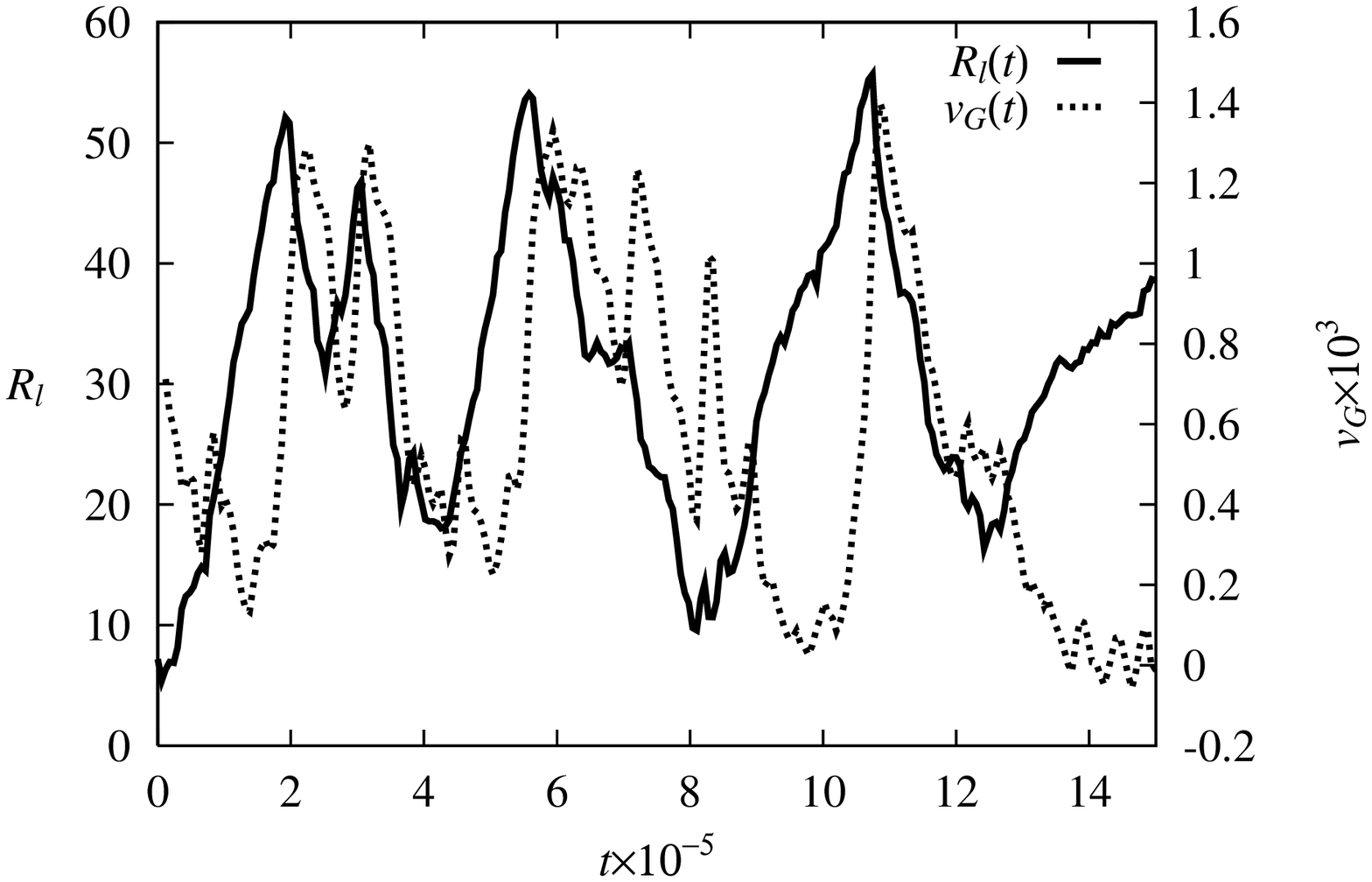}
\end{center}
%\end{figure}
\clearpage

\newpage
%\begin{figure}[t!]
\begin{center}
\resizebox{!}{7mm}{FIG.~6}\\\vspace*{25mm}
\leavevmode\epsfxsize=80mm
\epsfbox{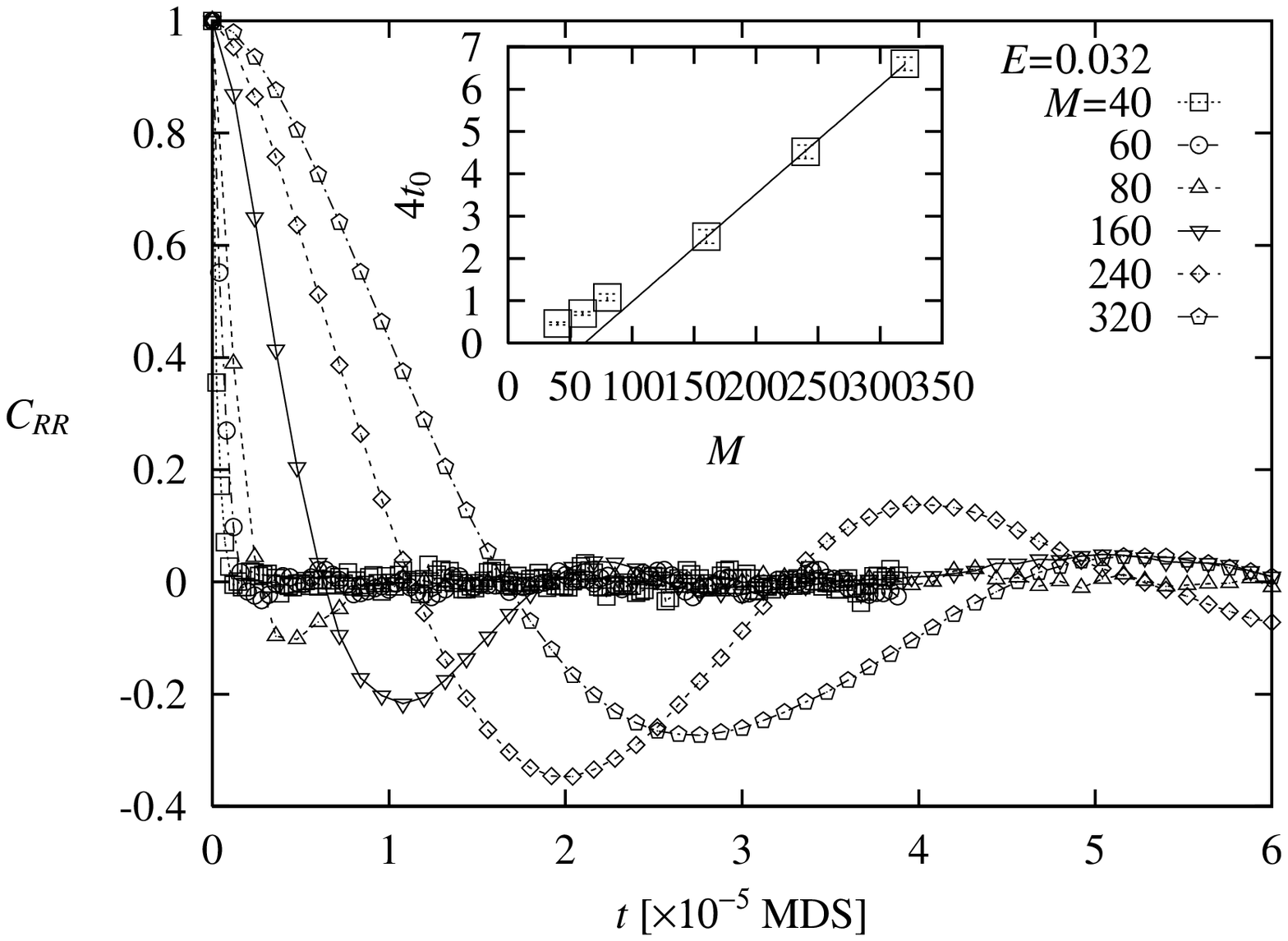}
\end{center}
%\end{figure}
\clearpage

\newpage
%\begin{figure}[t!]
\begin{center}
\resizebox{!}{7mm}{FIG.~7}\\\vspace*{25mm}
\leavevmode\epsfxsize=80mm
\epsfbox{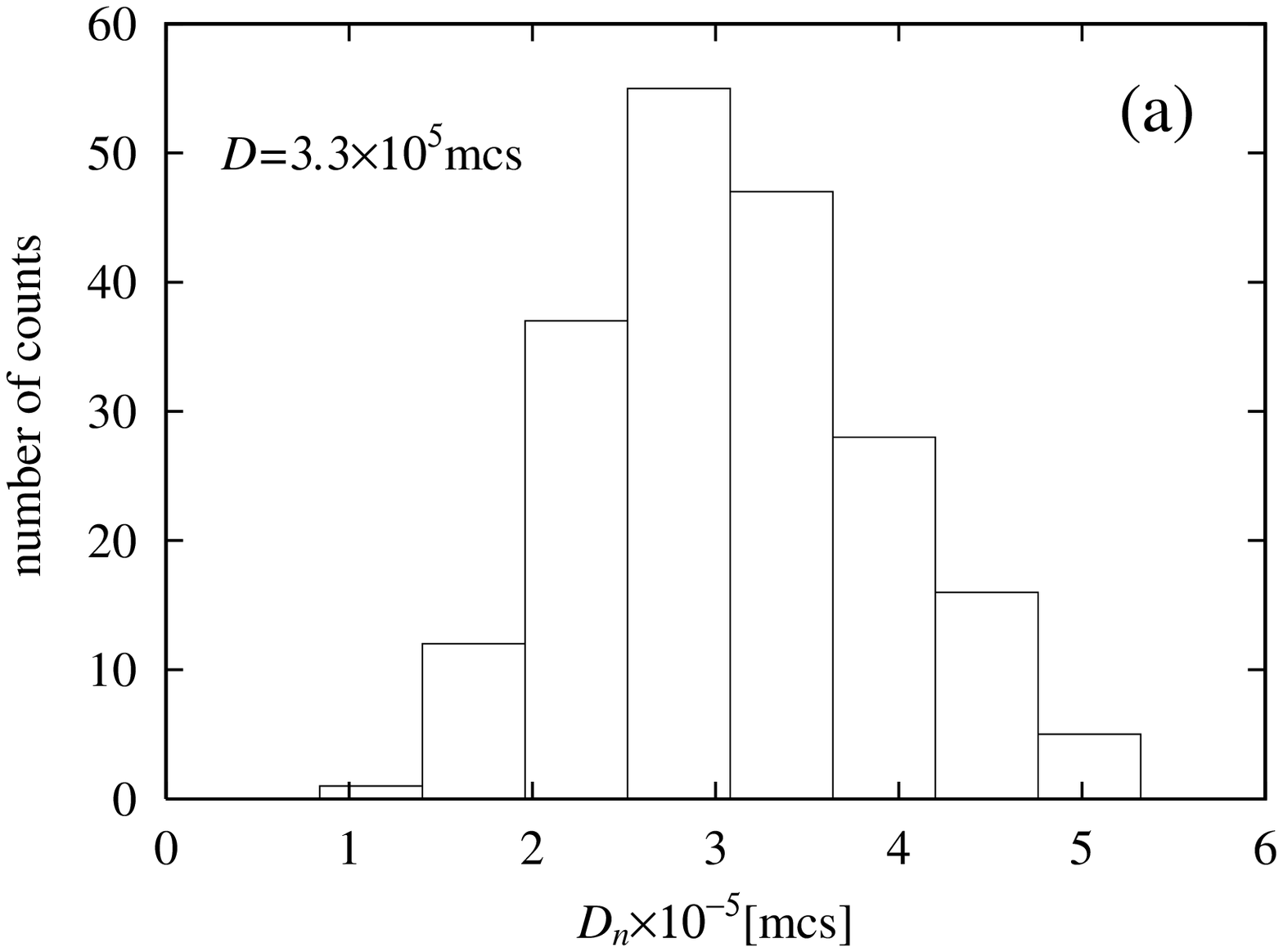}
\leavevmode\epsfxsize=80mm
\epsfbox{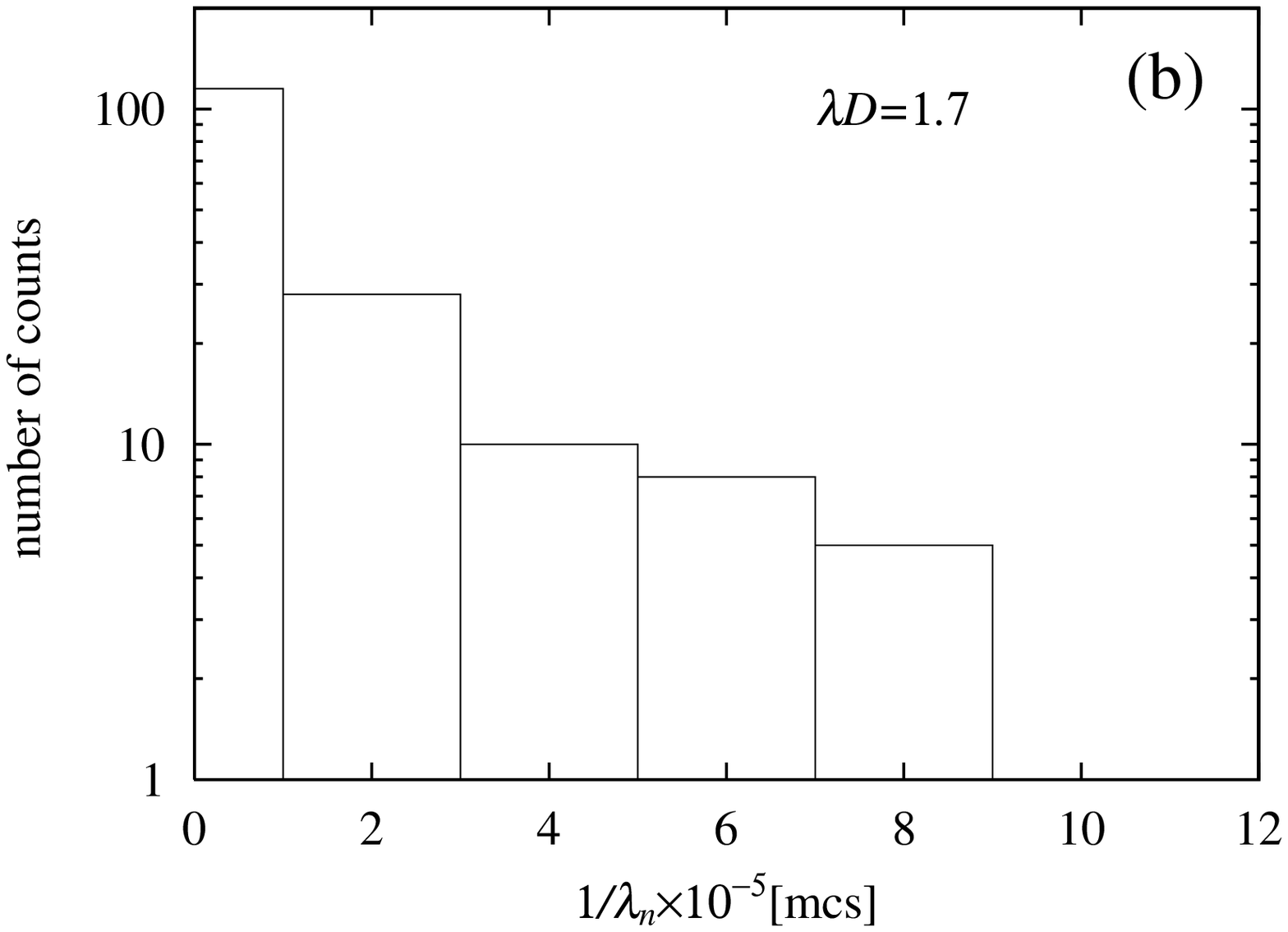}
\end{center}
%\end{figure}
\clearpage

\newpage
%\begin{figure}[t!]
\begin{center}
\resizebox{!}{7mm}{FIG.~8}\\\vspace*{25mm}
\leavevmode\epsfxsize=80mm
\epsfbox{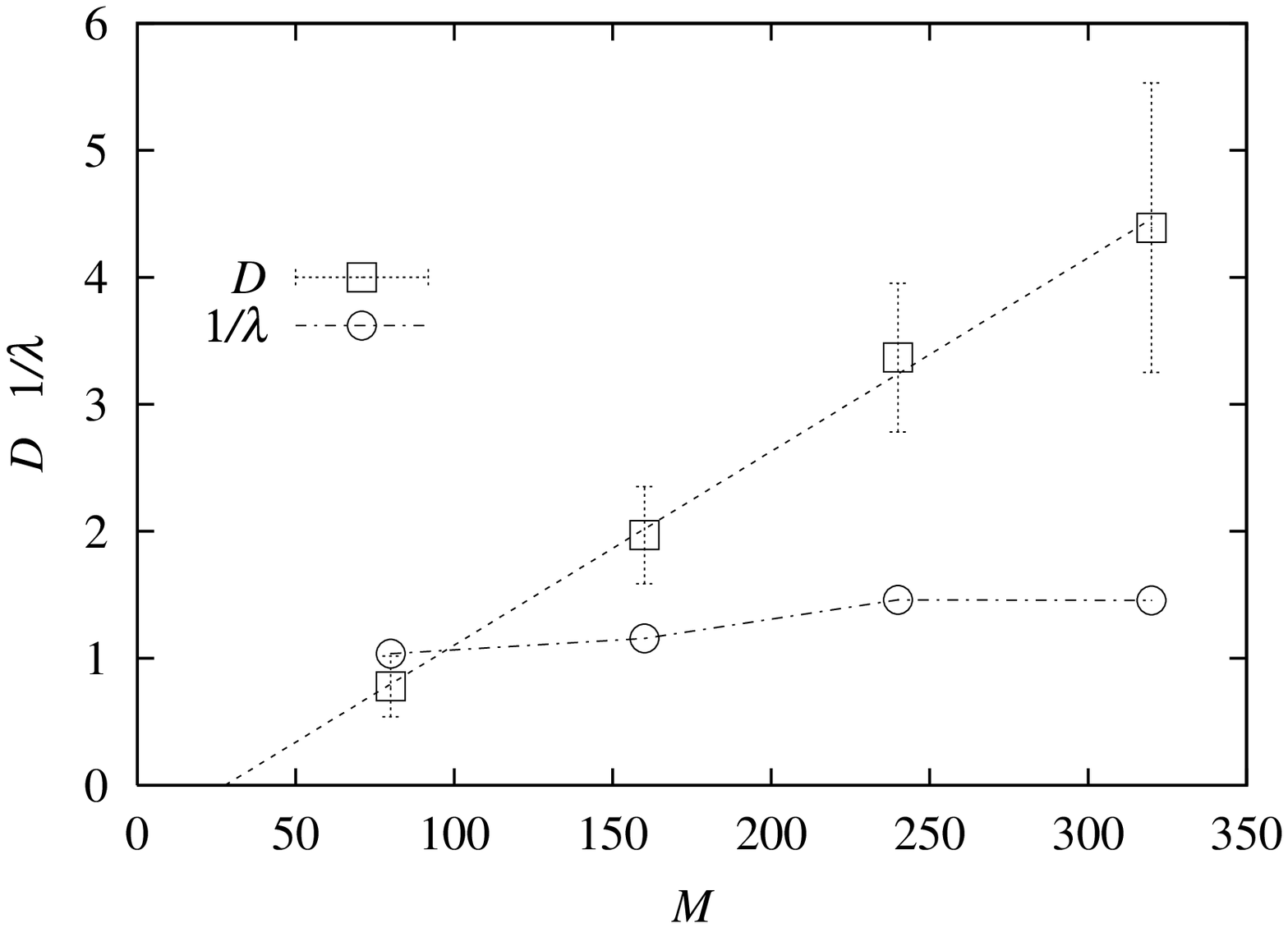}
\end{center}
%\end{figure}
\clearpage

\newpage
%\begin{figure}[t!]
\begin{center}
\resizebox{!}{7mm}{FIG.~9}\\\vspace*{25mm}
\leavevmode\epsfxsize=80mm
\epsfbox{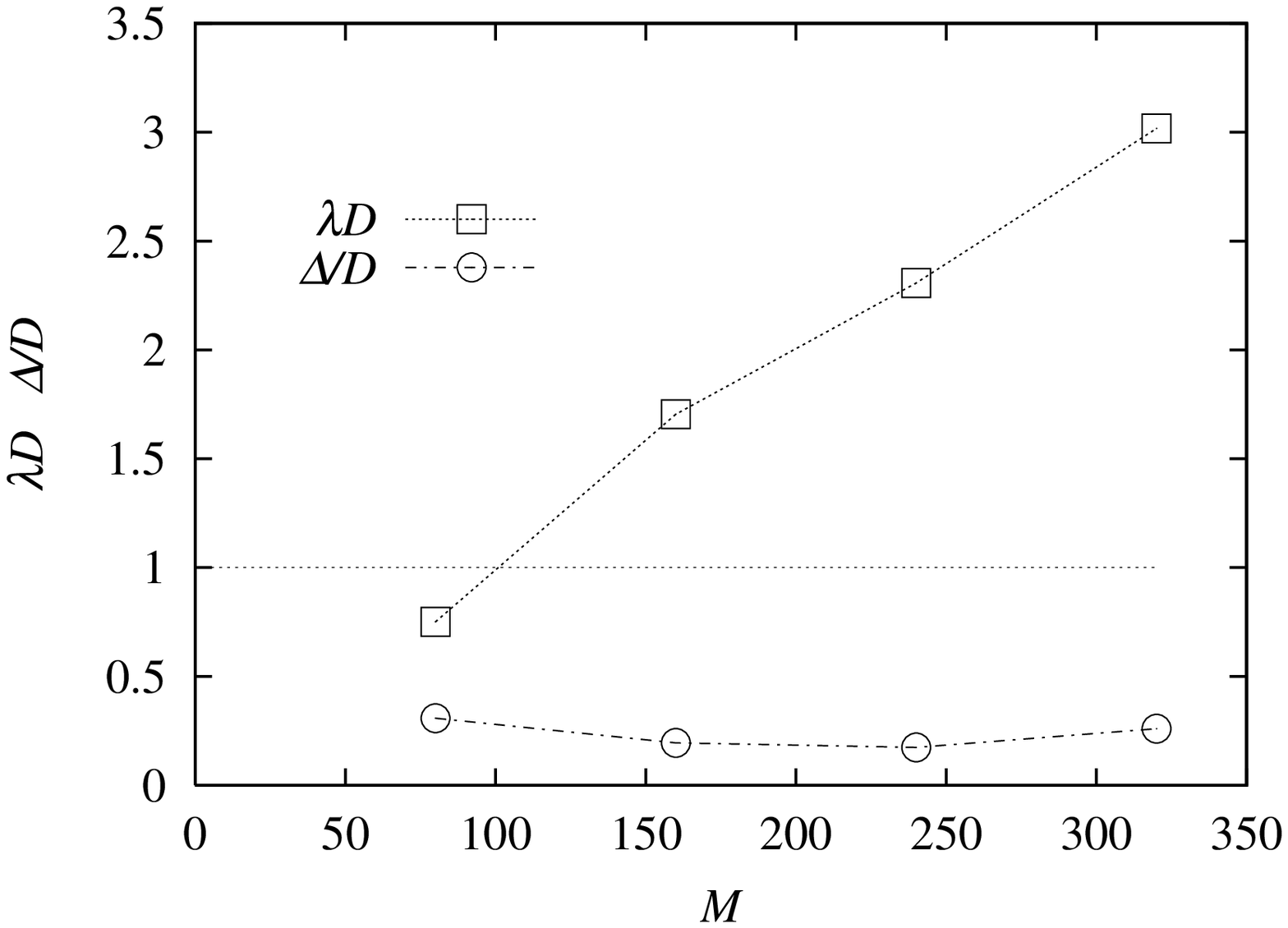}
\end{center}
%\end{figure}
\clearpage

\newpage
%\begin{figure}[t!]
\begin{center}
\resizebox{!}{7mm}{FIG.~10}\\\vspace*{25mm}
\leavevmode\epsfxsize=80mm
\epsfbox{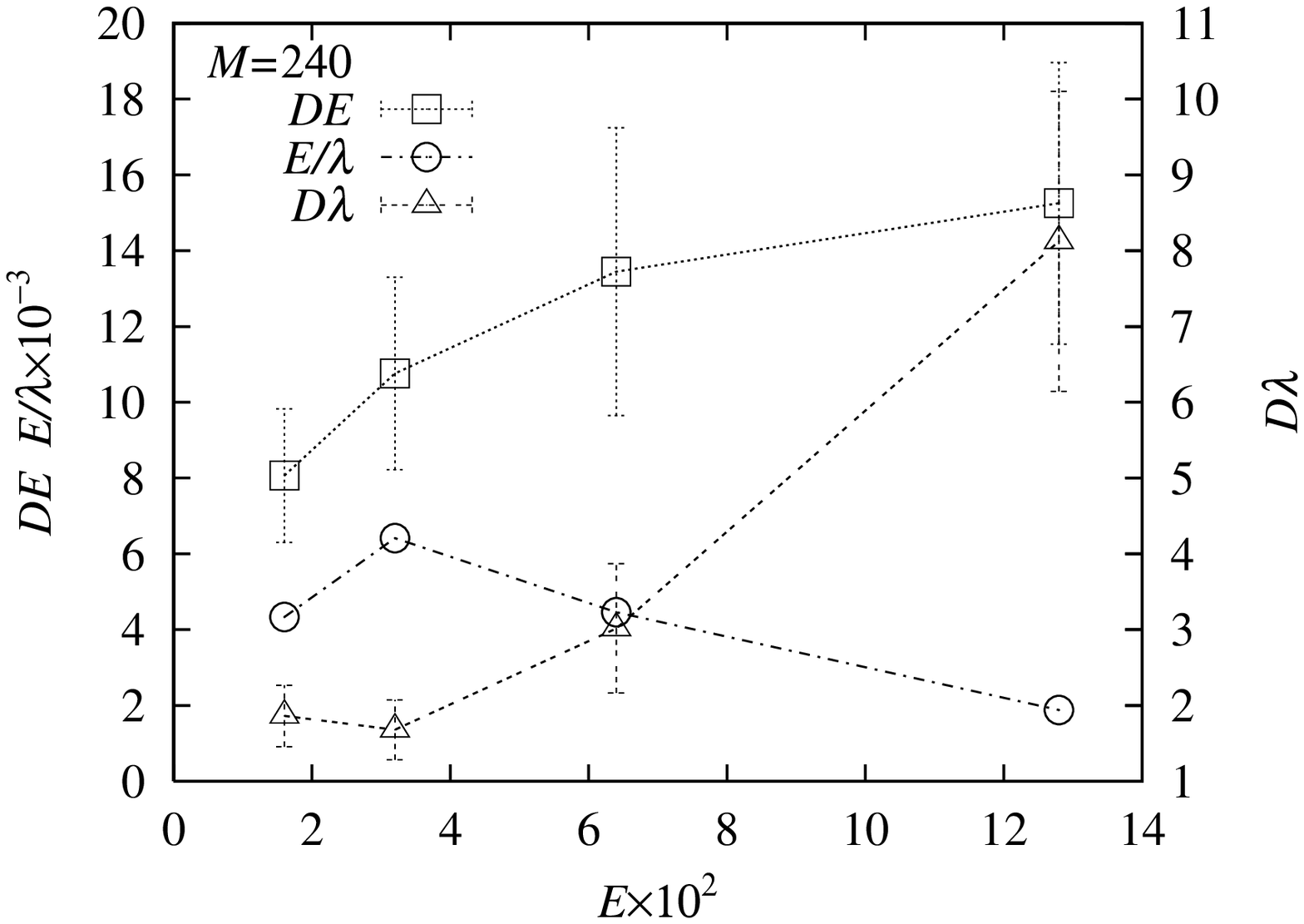}
\end{center}
%\end{figure}
\clearpage

\newpage
%\begin{figure}[t!]
\begin{center}
\resizebox{!}{7mm}{FIG.~11}\\\vspace*{25mm}
\leavevmode\epsfxsize=80mm
\epsfbox{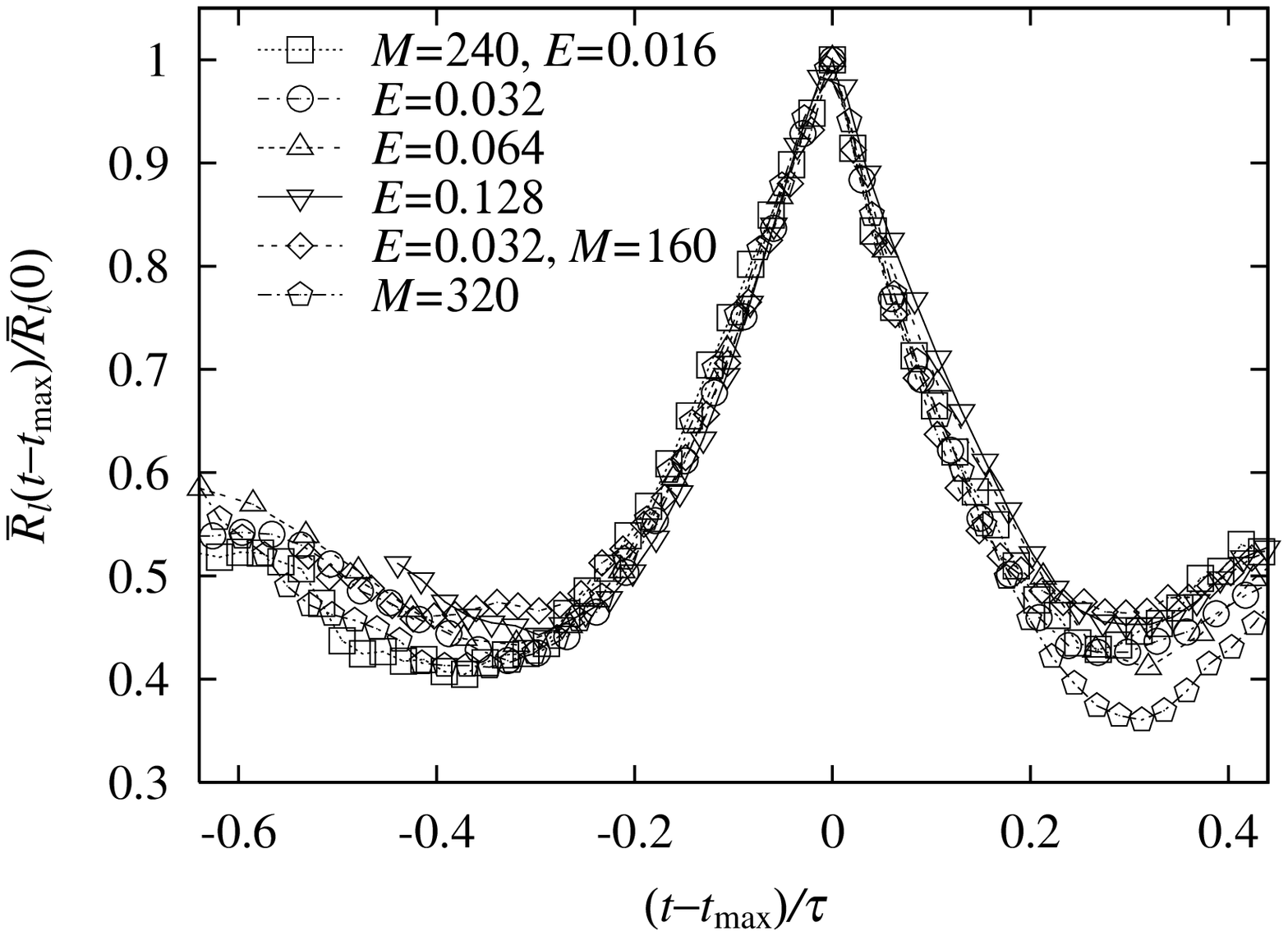}
\end{center}
%\end{figure}
\clearpage

\end{document}